\documentclass[aps,pra,superscriptaddress,twocolumn,floatfix,showpacs]{revtex4-1}
\usepackage{graphicx,color}
\usepackage[utf8]{inputenc}
\usepackage{amsmath,,amssymb,bm,amsfonts,dsfont,mathrsfs,amsthm}
\usepackage{braket}
\usepackage{txfonts,comment}
\usepackage[colorlinks=true,linkcolor=blue,citecolor=blue,urlcolor=blue]{hyperref}
\usepackage{soul}

\begin{document}

\title{Current Trends in Global Quantum Metrology}
\author{Chiranjib Mukhopadhyay} 
\affiliation{Institute of Fundamental and Frontier Sciences, University of Electronic Sciences and Technology of China, Chengdu 611731, China}
\affiliation{Key Laboratory of Quantum Physics and Photonic Quantum Information, Ministry of Education, University of Electronic Science and Technology of China, Chengdu 611731, China}

\author{Victor Montenegro} 
\affiliation{Institute of Fundamental and Frontier Sciences, University of Electronic Sciences and Technology of China, Chengdu 611731, China}
\affiliation{Key Laboratory of Quantum Physics and Photonic Quantum Information, Ministry of Education, University of Electronic Science and Technology of China, Chengdu 611731, China}

\author{Abolfazl Bayat} 
\email{abolfazl.bayat@uestc.edu.cn}
\affiliation{Institute of Fundamental and Frontier Sciences, University of Electronic Sciences and Technology of China, Chengdu 611731, China}
\affiliation{Key Laboratory of Quantum Physics and Photonic Quantum Information, Ministry of Education, University of Electronic Science and Technology of China, Chengdu 611731, China}

\begin{abstract}
Quantum sensors are now universally acknowledged as one of the most promising near-term quantum technologies. The traditional formulation of quantum sensing introduces a concrete bound on ultimate precision through the so-called local sensing framework, in which a significant knowledge of prior information about the unknown parameter value is implicitly assumed. Moreover, the framework provides a systematic approach for optimizing the sensing protocol. In contrast, the paradigm of global sensing aims to find a precision bound for parameter estimation in the absence of such prior information. In recent years, vigorous research has been pursued to describe the contours of global quantum estimation. Here, we review some of these emerging developments. These developments are both in the realm of finding ultimate precision bounds with respect to appropriate figures of merit in the global sensing paradigm, as well as in the search for algorithms that achieve these bounds. We categorize these developments into two largely mutually exclusive camps; one employing Bayesian updating and the other seeking to generalize the frequentist picture of local sensing towards the global paradigm. In the first approach, in order to achieve the best performance, one has to optimize the measurement settings adaptively. In the second approach, the measurement setting is fixed, however the challenge is to identify this fixed measurement optimally.
\end{abstract}

\maketitle

\tableofcontents

\section{Introduction}
\label{sec:Introduction}

Soon after the formulation of modern quantum mechanics, it was realized that the peculiarities of quantum systems, such as superposition and entanglement,  are extremely sensitive to external effects. This extreme sensitivity has paved the way for a new generation of \emph{quantum-enhanced} sensors whose precision theoretically surpasses any classical sensor~\cite{paris2009quantum,degen2017quantum,Braun-2018,montenegro2024review}. Such quantum systems can be miniaturized at the atomic level allowing for sensors to operate at nano-scales \cite{wu2022enhanced,chen2022immunomagnetic,du2024single,demille2024quantum}, as well as being capable of detecting extremely faint signals \cite{wang2024quantum,zhang2024quantum,sahoo2024enhanced,vetter2022zero}. Since its first inception~\cite{giovannetti2004quantum}, quantum sensing has now become a fast-maturing field with potential applications across various fields, including detection of gravitational waves in fundamental physics \cite{tse2019quantum}, searching for dark matter~\cite{jiang2021search,jiang2024long,wang2022limits,hartley2024quantum,huang2024new}, construction of atomic clocks with unprecedented precision~\cite{schulte2020prospects, giovannetti2001quantum,schleier2011cavity}, magnetic imaging in medical and microscopic settings~\footnote{See \cite{degen2017quantum} and references therein}, gravity estimation~\cite{qvarfort2018gravimetry, qvarfort2021optimal, rademacher2020quantum, armata2017quantum, gietka2019supersolid, kritsotakis2018optimal, montenegro2024heisenberg}, and gravity cartography for mineral prospecting~\cite{stray2022quantum}, and quantum radar~\cite{lanzagorta2012quantum,maccone2020quantum,karsa2024quantum}. In terms of platforms, quantum sensors have been envisaged in optical systems~\cite{dowling2015quantum},  nitrogen-vacancy centers~\cite{doherty2013nitrogen,rong2018searching,wang2019nanoscale,xie2020dissipative,rembold2020introduction,jiang2023quantum}, ion-traps~\cite{roos2006designer,wineland2011quantum,gilmore2021quantum}, neutral atom arrays~\cite{ding2022enhanced,liu2023electric,wu2023quantum}, superconducting qubits~\cite {behr2012development,wang2019heisenberg, shlyakhov2018quantum}, quantum dots~\cite{von2019quantum, gerster2018robust,chan2018assessment} or optomechanical setups \cite{schliesser2009resolved, xia2023entanglement,marti2024quantum,montenegro2022probing, montenegro2020mechanical}.

The program of quantum sensing relies on four crucial steps: \textit{(i)} probe preparation; \textit{(ii)} engineering probe evolution to encode the  unknown parameter of interest in the state of the probe; \textit{(iii)} performing a measurement on the probe; and \textit{(iv)} constructing an estimator which infers the unknown parameter from measurement outcome statistics. In order to get the best possible sensing performance, all the four steps need to be optimized. \textcolor{black}{In the first concrete proposal of quantum-enhanced sensing capability based on the Ramsey interferometry based phase estimation task ~\cite{giovannetti2001quantum,giovannetti2004quantum,giovannetti2006quantum}, the unknown parameter is simply encoded through a unitary phase shift}. The primary bottleneck is the probe preparation step, which requires preparing multiparty entangled Greenberger-Horne-Zeilinger (GHZ)-type states. The original GHZ-based quantum sensing scheme is limited to the class of sensing problems where the encoding of the unknown parameter \emph{can} be represented as a rotation and in any case, are exceptionally fragile to noise and particle loss. Interacting systems where quantum features emerge naturally in their spectrum offer a fundamentally different approach to quantum sensing~\cite{montenegro2024review}. Unlike GHZ-based quantum sensing, the entire spectrum carries information about the parameter thus making the probe more robust against imperfections. This is particularly evident in the long literature of criticality based many-body quantum sensors encompassing first-order~\cite{sarkar2024exponentially}, second-order~\cite{zanardi2006ground,garbe2020critical,mukhopadhyay2024modular}, topological~\cite{budich2020non,koch2022quantum,sarkar2022free,zhang2023topological,mukhopadhyay2024modular}, Floquet~\cite{lang2015dynamical,mishra2021driving,mishra2022integrable}, dissipative~\cite{chu2021dynamic,yang2019floquet}, and time-crystal~\cite{iemini2024floquet,shukla2024prethermal,montenegro2023boundary,cabot2024continuous,yousefjani2024discrete} phase transitions, see Ref.~\cite{montenegro2024review} for a detailed review. Instead, the primary bottleneck is shifted towards optimization of probe initialization, evolution, and measurement. Advances in quantum control theory has had immense impact on optimization of the probe evolution step~\cite{pang2017optimal,sekatski2017quantum,fallani2022learning}. Meanwhile, the emergence of noisy intermediate scale quantum (NISQ) devices allows simultaneous optimization of probe initialization and measurement steps~\cite{beckey2022variational,marciniak2022optimal,kaubruegger2019variational, meyer2021variational, ye2024essay}. 

The mathematical machinery behind quantum sensing is the extremely well-studied theory of statistical inference \cite{vantrees1968}. From this perspective, the role of quantum theory is to provide a probability model for which statistical inference theory can be applied. However, let us note that in the context of parameter estimation, quantum probabilities behave fundamentally differently than classical probabilities \cite{Holevo:1414149,parthasarathy2012introduction}. Thus, while in the absence of quantum features, the uncertainty of estimation scales at best as per the shot-noise limit (also known as standard limit) with respect to resources such as sensor size and sensing time, quantum properties may improve this scaling, a phenomenon known as \textit{quantum-enhanced sensitivity} \cite{giovannetti2001quantum, giovannetti2004quantum, giovannetti2006quantum}. However, to achieve this enhancement, one has to optimize the final measurement basis for any given probe initialization and interaction. Crucially, the optimal measurement itself may depend on the unknown parameter value, which seems paradoxical since it is precisely the unknown parameter value we want to estimate in the first place. Thus, the enhancement can only be achieved in two circumstances: (i) when the optimal measurement turns out to be independent of the unknown parameter value; and (ii) when we adopt the \emph{local} sensing paradigm, where a rough knowledge about the unknown parameter value is assumed beforehand allowing us to set up an approximately optimal measurement. Apart from some very specific problems, the optimal measurement strategy generically depends on the unknown parameter. This implies that, in the absence of prior information about the unknown parameter, the so-called \textit{global} sensing regime, a distinct formulation is required to quantify the precision, along with a systematic method for optimizing it.

{\color{black} From a practical perspective, global sensing becomes very important when the probe is expected to operate in an unknown territory. This can be the case for under-ocean and outer-space explorations where the parameters are unknown. An illustrative example is Pioneer 10, as the first spacecraft which successfully crossed the asteroid belt and reached the planet Jupiter in 1970s. The probe got seriously damaged  by unexpectedly high magnetic field of Jupiter. We now know that the magnetic field of Jupiter is $\sim 10^4$ times stronger than the earth, which was unknown at the time of Pioneer 10. The same story happened to the early Russian exploratory probes to Venus where the unexpectedly high temperature resulted in electronic malfunctioning. These examples clearly show that for space exploration our electronics and sensors must be capable of performing over a large operating window.    }

There are now several reviews on quantum sensing and metrology, covering the field from introductory perspectives \cite{paris2009quantum} to mathematically oriented treatments \cite{petz2011introduction,liu2020quantum,meyer2021fisher}, experimental and technology-focused approaches \cite{degen2017quantum,crawford2021quantum}, and specific classes of platforms, such as photonic \cite{pirandola2018advances} and many-body sensors \cite{montenegro2024review}. Nonetheless, there has been significant recent progress on global quantum sensing which have not made it so far to a coherent review. In this article, we address this issue by reviewing these recent developments for the benefit of current and future researchers. This review is organized in the following way. In Sec~\ref{sec:From Local to Global Sensing with Quantum Probes}, we discuss the general setting of global estimation problems. In Sec.~\ref{sec:Bayesian Strategies for Global Quantum Sensing}, we review global quantum sensing strategies employing Bayesian updating. In Sec.~\ref{sec:Frequentist Strategies for Global Quantum Sensing}, we complement them by reviewing the literature on global estimation from a frequentist standpoint. We finally conclude with a description of outstanding challenges and looking forward to future developments in the field in Sec.~\ref{sec:Outlook}.


\section{From Local to Global Sensing with Quantum Probes}
\label{sec:From Local to Global Sensing with Quantum Probes}
In the mid-1940s, H. Cram\'{e}r~\cite{cramer1946mathematical} and C. R. Rao~\cite{rao1992information} independently established a lower bound for the estimation of statistical parameters, known as the Cram\'{e}r-Rao inequality~\cite{paris2009quantum, rao1992information}:
\begin{equation}
\mathrm{MSE}(\hat{\theta}) \geq M^{-1}F_C^{-1}(\theta).\label{eq_MSE}
\end{equation}
On the left-hand side above, $\mathrm{MSE}(\hat{\theta})=\mathbb{E}_{x|\theta}\left[(\hat{\theta} - \theta)^2 \right]$ is the mean square error of $\hat{\theta}$, $\theta$ is the unknown parameter to be estimated, $\hat{\theta}(x):=\hat{\theta}$ is the estimator function that maps the measurement data $x$ (here, without loss of generality, assumed to be continuous) to parameter space, and $\mathbb{E}_{x|\theta}[\square]=\int_\mathbb{R} p(x|\theta) \square dx$ is the expectation value with respect to the probability distribution $p(x|\theta)$. Refer to R. A. Fisher~\cite{fisher1920amathematical} for determining the accuracy of observations using both the mean error and the mean square error. On the right-hand side of Eq.~\eqref{eq_MSE}, $M$ is the number of measurement trials and 
\begin{equation}
    F_C(\theta)=\int_\mathbb{R} dx p(x|\theta)\left(\frac{\partial \mathrm{ln}p(x|\theta)}{\partial\theta}\right)^2
\end{equation}
is the classical Fisher information (CFI)~\cite{fisher1925theory}. This CFI function quantifies the maximum amount of information extractable about the parameter $\theta$ for a specific measurement. Assuming the use of unbiased estimators---those for which the expectation value of the estimator function converges to the actual unknown parameter $\theta$ in the asymptotic limit of measurements, that is $E_{x|\theta}[\hat{\theta}]=\theta$---the mean square error simplifies to the variance of the estimator for $\theta$, i.e. $\mathrm{Var}[\hat{\theta}]$. Consequently, the Cram\'{e}r-Rao inequality becomes:
\begin{equation}
\mathrm{Var}[\hat{\theta}] \geq M^{-1}F_C^{-1}(\theta).\label{eq_CRB}
\end{equation}
The inequality in Eq.~\eqref{eq_CRB} depends solely on the observed data $x$. Eq.~\eqref{eq_CRB} reveals a deeper connection between CFI and the asymptotic efficiency of an estimator. This relationship was rigorously formalized by H\'{a}jek-Le Cam in their convolution theorem~\cite{LeCam-1986, Hajek1970} (see also Inagaki~\cite{Inagaki1970}), which provides a broader framework for understanding the asymptotic behavior of estimators. Specifically, an estimator is considered efficient if its limiting distribution is purely normal with variance $ M^{-1} F_C^{-1}(\theta)$. Conversely, if the estimator is inefficient, additional noise appears as a convolution in its limiting distribution. It has been proven that, under certain regularity conditions, the maximum-likelihood estimator~\cite{fisher1912onanabsolute, fisher1922onthemathematical, fisher1925theory} is asymptotically efficient~\cite{borovkov1999mathematical}. Assuming that $\ln p(x|\theta)$ is twice differentiable, the CFI can also be expressed as $F_C(\theta) = -\int_\mathbb{R} p(x|\theta) \frac{\partial^2 \ln p(x|\theta)}{\partial \theta^2} dx$~\cite{fisher1925theory}. C. W. Helstrom~\cite{helstrom1968theminimumvariance, helstrom1967minimum} proposed that by incorporating principles from quantum theory, one can derive a tighter lower bound than that provided by the Cram\'{e}r-Rao inequality, see also A. Holevo~\cite{Holevo:1414149}. 

In quantum mechanics, probability distributions are operationally defined by the Born rule: $p(x|\theta) = \mathrm{Tr}[\Pi_x \rho(\theta)]$. In this expression, $\mathrm{Tr}[\cdot]$ denotes the trace operation, $\rho(\theta)$ is the quantum state that encodes the unknown parameter $\theta$, and $\Pi_x$ is a positive operator-valued measure (POVM) that satisfies $\int \Pi_xdx = 1$ and $\Pi_x \geq 0$~\cite{janos2021quantum}. By applying the Born rule in the CFI, maximizing the CFI over all POVMs, and using the Cauchy–Schwarz inequality, one can derive a fundamental lower bound for the uncertainty of an unknown parameter known as the quantum Cram\'{e}r-Rao inequality~\cite{helstrom1968theminimumvariance, helstrom1967minimum, paris2009quantum}:
\begin{equation}
\mathrm{Var}[\hat{\theta}] \geq M^{-1}F_C^{-1}(\theta)\geq M^{-1}F_Q^{-1}(\theta),\label{eq_QCRB}
\end{equation}
where $F_Q(\theta)$ is known as the quantum Fisher information (QFI), and by definition, it simply corresponds to the POVM that maximizes the CFI, i.e., $F_Q(\theta) := \max\limits_{\{\Pi_x\}} F_C(\theta)$. Eq.~\eqref{eq_QCRB} defines the ultimate precision limits allowed by quantum mechanics and represents the cornerstone of quantum parameter estimation. {\color{black}  In practice, one does not need to perform a brute force maximization over all possible measurements. In  fact, it has been shown that the optimal measurements lie along the eigenbases of the so-called \emph{symmetric log-derivative} (SLD) operator $\mathcal{L}$ defined as
\begin{equation}
    \frac{\partial \rho}{\partial \theta} = \frac{\mathcal{L}\rho + \rho \mathcal{L}}{2}.
\end{equation}
The corresponding QFI is then given by $F_Q=\mathrm{Tr}\left[ \rho(\theta) \mathcal{L}^2\right]$.}

After performing the measurement, either optimal or not, classical techniques are then used to process the outcome and estimate the parameter. The goal is to get as close as possible to the QFI bound~\cite{paris2009quantum, giovannetti2004quantum, giovannetti2006quantum, abiuso2024fundamental, puig2024dynamical}. Both CFI and QFI have been studied from a geometric perspective to distinguish probability distributions or quantum states in the infinitesimal vicinity of $\theta$~\cite{braunstein1994statistical}, see J. S. Sidhu et al.~\cite{sidhu2020geometric} for a thorough review on the geometry aspect of quantum parameter estimation. In Ref.~\cite{meyer2021fisher}, the relevance of CFI and QFI for near-term quantum devices have been reviewed. Extensive studies have explored generalizations to multiparameter scenarios. See, for instance, Liu et al.~\cite{liu2020quantum} and F. Albarelli et al.~\cite{albarelli2020perspective} for a more detailed discussion.
  
\begin{figure}[t]
    \centering
    \includegraphics[width=\linewidth]{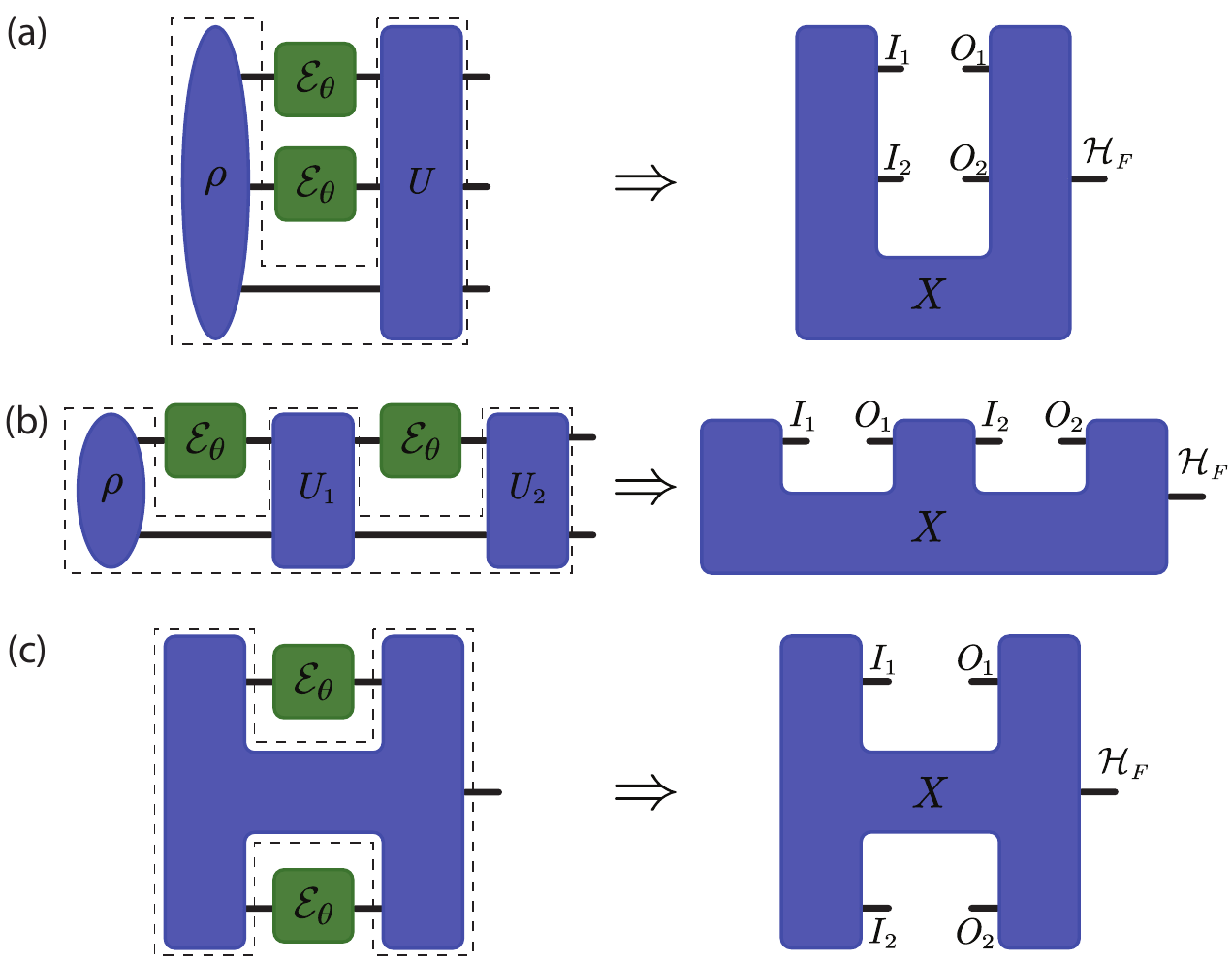}
    \caption{\textbf{Schematic of quantum sensing strategies.} (a) parallel strategies, (b) sequential strategies, and (c) general indefinite-causal-order strategies. The left column illustrates that a strategy is an arrangement of physical operations such as initial-state preparations and adaptive controls. The right column depicts that a strategy amounts to a supermap which, akin to completely positive maps, can be described by a positive-semidefinite operator $X$. Figure taken from Ref.~\cite{zhou2024strict}.}
    \label{fig_hierarchy}
\end{figure}

The entire discussion above, concerning the inference of unknown parameters using the quantum Cram\'{e}r-Rao bound, falls under the framework of \textit{local} estimation theory~\cite{helstrom1967minimum, braunstein1994statistical}. In this context, local refers to minimizing the variance of the estimator by maximizing the CFI at a specific value of the unknown parameter~\cite{paris2009quantum}, with a general classification of four distinct quantum sensing strategies can be made as follows: (i) parallel strategies~\cite{giovannetti2006quantum}; (ii) sequential strategies~\cite{giovannetti2006quantum}; (iii) causal superposition strategies~\cite{mukhopadhyay2018superposition,zhao2020quantum,mothe2024reassessing}; and (iv) general indefinite-causal-order strategies~\cite{liu2023optimal,caleffi2023beyond}, see Fig.~\ref{fig_hierarchy}.
Despite the elegance and power of local quantum estimation theory, several important subtleties must be addressed: 
\begin{enumerate}
    \item[(I)] \textbf{Unbiasedness:} The Cram\'{e}r-Rao inequality is meaningful only to unbiased estimators, yet many practical estimators are biased.
    \item[(II)] \textbf{Regularity conditions:} CFI assumes regularity conditions, conditions that are often difficult to verify in practice.
    \item[(III)] \textbf{Limited data:} While asymptotically efficient estimators estimators exist, selecting an efficient estimator when the available measurement data is limited is challenging.
    \item[(IV)] \textbf{Global sensing:} The optimal POVM that maximizes the CFI typically depends on the unknown parameter $\theta$.
\end{enumerate}

These subtleties can be addressed as follows: Subtlety (I) concerns the unbiasedness of the estimator. In practice, determining if an estimator is unbiased is difficult. This is because unbiasedness can only be reliably assessed with large sample sizes, where one can test whether the expectation value of observed data aligns with the true parameter value~\cite{bj/1186078362}. The second subtlety (II) involves ensuring regularity conditions, such as the existence of integrals and the differentiability of probability distributions~\cite{bj/1186078362}. This can be addressed by constructing probability distributions based on specific families of statistical models or a given classical or quantum statistical model~\cite{borovkov1999mathematical}. The third subtlety (III)~\cite{fisher1912onanabsolute} can be addressed by identifying the measurement scheme that achieves the optimal single-shot mean square error~\cite{personick1971application} and applying this measurement in repeated experiments~\cite{rubio2019quantum, rubio2020bayesian,rubio2024first}. Alternative approaches employ adaptive strategies, which often require multiple complex measurement types~\cite{Braun-2018, Mehboudi-2016}—see also Refs.~\cite{Okamoto-2012, Okamoto-2017, Fujiwara-2006, Fujiwara-2011, Wiseman-1995, Higgins-2007, Armen-2002} for examples of adaptive quantum state estimation procedures using photons. The final subtlety (IV) involves saturating Eq.~\eqref{eq_QCRB} by selecting an optimal POVM. However, this choice of POVM depends on the unknown parameter, requiring significant prior knowledge of the parameter $\theta$ to reach the fundamental uncertainty limit~\cite{montenegro2021global}. {\color{black} As a minimal qualitative example, let us consider a single qubit quantum probe which is given by the state
\begin{equation}\label{eq:probe_1qubit}
    \rho(\theta)=p\ket{\psi(\theta)}\bra{\psi(\theta)}+\frac{1-p}{2} \mathbb{I}
\end{equation}
where, $\ket{\psi(\theta)}{=}\cos(\theta/2)\ket{0}{+}\sin(\theta/2)\ket{1}$,  $\mathbb{I}$ is the $2{\times} 2$ identity matrix and $p$ is the degree of the purity of the probe. The goal is to estimate the unknown parameter $\theta$. If one uses $z$-measurement, described by the projectors $\Pi_0{=}\ket{0}\bra{0}$ and $\Pi_1{=}\ket{1}\bra{1}$, the corresponding CFI is given by $F_C{=}p^2\sin^2(\theta){/}(1-p^2\cos^2(\theta))$, which is maximal for $\theta=\pi/2$, irrespective of $p$. On the other hand, if one uses the $x$-measurement, described by the projectors $\Pi_+{=}\ket{+}\bra{+}$ and $\Pi_-{=}\ket{-}\bra{-}$ with $\ket{\pm}{=}(\ket{0}{\pm}\ket{1})/\sqrt(2)$, the corresponding CFI becomes $F_C{=}p^2\cos^2(\theta){/}(1-p^2\sin^2(\theta))$, which is maximal for $\theta{=}0$. In general, one can show that for the probe in Eq.~(\ref{eq:probe_1qubit}) the SLD operator is given by
\begin{equation}
    \mathcal{L}=p \begin{bmatrix}
        -\sin \theta & \cos \theta \\
        \cos \theta & \sin \theta
    \end{bmatrix},
\end{equation}
whose eigenbases clearly depend on the unknown parameter $\theta$. This simple example implies that the optimal measurement indeed depends on the unknown parameter $\theta$. Therefore, without  complete prior information about $\theta$, it is not clear what measurement setting needs to be adopted. To address this issue, one has to formulate a consistent theory for global sensing which is the subject of this review article. }

The issue of global sensing was originally formulated and solved by H. L. Van Trees~\cite{vantrees1968} which looks for the POVM minimizing a suitable cost functional, averaged over all possible values of the parameter to be estimated~\cite{paris2009quantum}. \textcolor{black}{In terms of operational strategies, research into  quantum global estimation theory can largely be divided into two approaches. The first one is the so-called Bayesian approach, where the initial prior is sharpened over subsequent rounds by applying Bayes theorem and using the posterior probabilities for each round as priors for the next one. The second approach is the frequentist one, where the prior is not updated at each round and instead the focus is on minimizing some pre-decided figure of merit calculated by averaging the precision of local estimation with suitable weights determined by the prior. } 

\section{Bayesian Strategies for Global Quantum Sensing}
\label{sec:Bayesian Strategies for Global Quantum Sensing}
 Assuming a minimal  prior knowledge about the parameter given by a probability distribution function $z(\theta)$, the van Trees inequality considers that the average variance is given by the Bayesian Cram\'{e}r-Rao bound or posterior Cram\'{e}r-Rao bound~\cite{vantrees1968}:
\begin{equation}
\overline{\mathrm{Var}[\hat{\theta}]}=\int z(\theta) \left[\hat{\theta}(x) - \theta \right]^2 dxd\theta \geq \frac{1}{Z_F},  
\end{equation}
where $\hat{\theta}(x)$ is the estimator of $\theta$ (mapped from the measurement outcomes $x$), and $Z_F{=}M\int dxd\theta p(x{,}\theta)(\partial_\theta\log p(x{,}\theta))^2$ with $p(x{,}\theta){=}p(x{|}\theta)z(\theta)$ can be demonstrated to be~\cite{paris2009quantum}
\begin{equation}
    Z_F = M \int z(\theta)F_C(\theta)d\theta + M\int z(\theta)[\partial_\theta \mathrm{log}z(\theta)]^2 d\theta,
\end{equation}

where the first term accounts for the average of the CFI over the prior distribution $z(\theta)$ and the second term is the CFI of the prior distribution itself~\cite{vantrees1968}. See M. Fau$\ss$ et al.,~\cite{michael2021avariational} for variational proofs of both the classic and the Bayesian Cram\'{e}r-Rao bounds. A straightforward generalization of the quantum van Trees inequality can be derived~\cite{paris2009quantum, martinez2017quantum, suzuki2023bayesiannagaokahayashiboundmultiparameter}:
\begin{equation}
    \overline{\mathrm{Var}[\hat{\theta}]} \geq \frac{1}{Z_F} \geq \frac{1}{Z_Q},
\end{equation}
where $Z_Q{=}M \int z(\theta)F_Q(\theta)d\theta + M\int z(\theta)[\partial_\theta \mathrm{log}z(\theta)]^2 d\theta$~\cite{suzuki2023bayesiannagaokahayashiboundmultiparameter}. 
Moreover, it has been stated that the van Trees inequality is not only a Bayesian analog of the Cram\'{e}r–Rao bound, but an instance of the Cram\'{e}r–Rao bound for a suitably chosen location model~\cite{elisabeth2024vantreesinequalityspirit}. In Ref.~\cite{jupp2010avantrees}, the van Trees inequality has been extended to the setting of smooth loss functions on manifolds, addressed in terms of differential geometry. Recently, K. Takatsu et al.,~\cite{takatsu2024generalizedvantreesinequality} provided a new non-asymptotic minimax lower bounds under minimal regularity assumptions, without requiring differentiability of functionals or regularity of statistical models (see the drawbacks of local sensing strategies mentioned above). 

In the Bayesian approach, the unknown parameter to be estimated $\theta$ is treated as a random variable. Before the estimation process begins, one explicitly defines a \textit{prior} distribution for this parameter, which represents any initial knowledge or assumptions about $\theta$. This prior distribution reflects what is known about $\theta$ before any new data or measurements are taken into account~\cite{rafal2015quantum}. The optimal strategy for the Bayesian approach is guaranteed by the optimality of covariant measurements in estimation problems that exhibit certain group symmetries~\cite{rafal2015quantum, Holevo:1414149}. Furthermore, the optimal estimation strategy is directly determined by the prior probability distribution that is assumed~\cite{rafal2015quantum}. While Bayesian approaches have been proposed in the spirit of local estimation, namely by minimizing the MSE~\cite{personick1971application}, the choice of a cost function is not limited to the MSE. In fact, any other cost function can be used, depending on the specific requirements or goals of the estimation. In general, one can consider 
\begin{equation}
    \langle C \rangle = \int d\theta d\bm{x}z(\theta)p(\bm{x}|\theta)C(\hat{\theta}(\bm{x}),\theta),
\end{equation}
where $C(\hat{\theta}(\bm{x}),\theta)$ is a general cost function which depends on the unknown parameter $\theta$ and its estimator $\hat{\theta}(\bm{x})$~\cite{rafal2015quantum}.

The flexibility in choosing the cost function allows for tailoring the strategy to different practical needs. For instance, in a phase estimation strategy with prior knowledge about the phase,
a suitable choice for a prior distribution is the flat distribution 
$z(\theta)=1/(2\pi)$ ($\theta$ as the unknown phase) with a cost function $C(\hat{\theta}(\bm{x}),\theta) = 4\sin^2[(\hat{\theta}(\bm{x}) - \theta] / 2)$. For a probe initialized in the  N00N states, i.e. $|\text{N00N}\rangle = 1/\sqrt{2}(|\text{N,0}\rangle + |\text{0,N}\rangle)$~\cite{dowling2008quantum} (with $N$ being the number of excitations), a unitary evolution leads to the Heisenberg limit of precision $\langle C \rangle \approx (\pi/N)^2$~\cite{berry2000optimal, luis1996optimum, summy1990phase}.
Note that, while N00N states are known to be the optimal state in the context of local phase estimation using QFI, the Bayesian framework leads to an entirely different optimal initial state (the \textit{sine} state), namely:
\begin{equation}
    |\psi\rangle=\sum_{n=0}^N \sqrt{\frac{2}{2+N}}\sin\left( \frac{n+1}{N+2}\pi \right)|n,N-n\rangle,\label{eq_sine_state}
\end{equation}
for reaching the Heisenberg limit~\cite{rafal2015quantum}. Indeed, in the asymptotic limit of large excitations $N$, the state in Eq.~\eqref{eq_sine_state} results in $\langle C \rangle \approx (\pi/N)^2$, where the cost function is defined as $C(\hat{\theta}(\bm{x}), \theta) = 4 \sin^2 \big( [\hat{\theta}(\bm{x}) - \theta] / 2 \big)$.
It is also possible to find global and local sensing methodologies to convey the same precision uncertainty in the asymptotic limit. For instance, in the estimation of a transmission coefficient, both Bayesian estimation methods and those based on QFI yield the same result when evaluated asymptotically, meaning that the estimators converge to the same precision as the amount of data increases~\cite{vaart1998asymptotic}. This also holds true in non-unitary dynamics~\cite{alipour2014quantum}. Indeed, asymptotic precision bounds have been established for specific cases, such as global dephasing~\cite{jarzyna2015true}, where the precision bound is found to match that given by the QFI (local sensing), and photonic losses~\cite{kolodynski2010phase}, where quantum enhancement provides a multiplicative improvement over classical strategies (shot noise limit). Based on the Bayesian framework above~\cite{personick1971application},  further applications including thermometry~\cite{rubio2021global, glatthard2022optimal,mehboudi2022fundamental, rodriguez2024strongly,abiuso2024optimal,jorgensen2022bayesian, boeyens2023probe} and critical sensing in many-body systems through feedback control \cite{salvia2023critical} have been formulated. 



Due to the current lack of tools for effectively evaluating the performance of global estimation strategies~\cite{meyer2023quantummetrologyfinitesampleregime, bavaresco2024designing}, Zhou et al.~\cite{zhou2024strict} have addressed this gap by developing a strict hierarchy for general global estimation strategies. This hierarchy, which is based on the general classification shown in Fig.~\ref{fig_hierarchy}, was established using an innovative method referred to as virtual imaginary-time evolution.
In their work, the authors demonstrate that it is possible to establish an equality between the information gained about a parameter in a global estimation process and the QFI associated with a corresponding virtual local estimation. In concreteness, it has been defined~\cite{zhou2020quantum}:
\begin{equation}
    \mathcal{J}=\mathrm{Tr}[\bar{\rho}S^2],
\end{equation}
where $\bar{\rho}=\int z(\theta)\rho(\theta)d\theta$ and $S$ obeys the implicit equation $2\overline{\theta\rho}=\bar{\rho}S{+}S\bar{\rho}$ with $\overline{\theta\rho}=\int z(\theta)\theta \rho(\theta)d\theta$.
In terms of the global estimation strategies above, Ref.~\cite{zhou2020quantum} establishes the following global sensing hierarchy as:
\begin{equation}
    \mathcal{J}_\mathrm{max}^{(i)} \leq \mathcal{J}_\mathrm{max}^{(ii)} \leq \mathcal{J}_\mathrm{max}^{(iii)} \leq \mathcal{J}_\mathrm{max}^{(iv)},
\end{equation} 
where superscripts (i) to (iv) represent the hierarchies enumerated in Fig.~\ref{fig_hierarchy}, $J_\mathrm{max}^{(k)}$ is the maximum of $\mathcal{J}$ over all the strategies of type $k$~\cite{zhou2024strict}; and where there is always a quantum parameter estimation for which $\mathcal{J}_\mathrm{max}^{(i)} < \mathcal{J}_\mathrm{max}^{(ii)} < \mathcal{J}_\mathrm{max}^{(iii)} < \mathcal{J}_\mathrm{max}^{(iv)}$~\cite{zhou2024strict}.

\textcolor{black}{At this point, let us remember that while the Cramer-Rao bound is asymptotically tight, other theoretical bounds recently developed may be better suited to real experimental conditions involving finite shots of measurement \citep{tsang2012ziv,chang2022evaluating,berry2015quantum,liu2016valid, aharon2023asymptoticallytightbayesiancramerrao}. Extension of these bounds to the global sensing scenario is likely to be extremely fruitful for practical pursposes.}

\subsection{Unknown Scale Global Sensing}

Our choice of prior usually inadvertently reveals some subtle foreknowledge or assumption about the parameter. As an example, consider the thermometry problem. If we choose a prior which is uniform, i.e., conveying maximal ignorance about the parameter value, between two distinct temperatures $T_{\min}$ and $T_{\max}$, that still assumes we roughly know the scale of the temperature. For example, if $T_{\min}$ and $T_{\max}$ are respectively $10^{-9}$K and $10^{3}$K, then choosing an uniform prior in this interval would mean contribution from lower temperatures are washed out by much bigger contributions from higher temperatures, so the optimal global sensing protocol may not be relevant at all for temperatures spanning several orders of magnitude. \textcolor{black}{This is relevant for many other parameters as well. For example, the resistance of electrical components may vary over  at least 9 orders of magnitude ranging from conductors to insulators. In these setups, the precision that the probe requires to provide varies with the scale of the parameter. For instance, mili Kelvin precision, i.e. $\delta T{\sim} 10^{-3}$, for measuring a temperature which is hundreds of Kelvin, i.e. $T{\sim} 10^2$K, can be considered as an exceptionally high precision resulting in signal-to-noise ratio of $T/\delta T {\sim} 10^5$. However, the same precision for measuring micro Kelvin temperatures, i.e. $T{\sim} 10^{-6}$K, results in a very poor signal-to-noise of $T/\delta T {\sim} 10^{-3}$. This shows that when the range of the unknown parameter varies over several scales one has to take this issue into careful consideration. }

\subsubsection{Unknown Scale Global Thermometry}

Let us assume we are not aware of the scale of the unknown temperature $T$, but we can access energy measurement data $E$. The authors of Ref.~\cite{rubio2021global} argue that for a hypothesis $\theta$ for the actual temperature, the conditional probabilities $p(E|\theta)$ must solely be a function of the dimensionless quantity $E/k_B\theta$, where $k_B$ is the Boltzmann constant. Now, suppose the energy data is scaled by some unknown constant $\gamma$, i.e., $E{\rightarrow}\gamma E$, and likewise hypothesis $\theta{\rightarrow}\gamma\theta$. If the problem is scale-invariant, then this should not offer us any new information under a reasonable choice of prior $z(\theta)$. Thus,  $z(\gamma \theta)/z(\theta) = 1/\gamma$, or equivalently, $z(\theta)\propto 1/\theta$. The next task is to build an estimator $\hat{\theta}(E)$ and a suitable \emph{deviation function} $\mathcal{D}[\hat{\theta}(E), \theta]$, whose minimization would maximize the precision of the probe in a local estimation scenario. Since the true temperature itself is unknown, the deviation function involves global minimization of this average uncertainty, i.e., the cost function becomes 

\begin{equation}
    {\bar{\epsilon}}_{\mathcal{D}} = \int dE d\theta p(E,\theta) \mathcal{D}[\hat{\theta}(E), \theta].
\end{equation}

However, unlike the approach in the previous sections, this deviation function is no longer simply the $l_2$-distance between the hypothesis and the estimate, since this destroys the scale-invariance. Instead Ref.~\cite{rubio2021global} maps this problem into one with dimensionless hypothesis $x = \alpha \log(k_B \theta/\epsilon_0)$ to explicitly maintain scale invariance and then they consider $\mathcal{D}[\hat{x}(E), x] = |x(E) - x|^k$ deviation. Note that $\epsilon_0$ is an arbitrary constant with dimensions of energy which defines the scale. Mapping back to the original problem, this leads to the following family of deviation functions satisfying the scale-invariance condition 

\begin{equation}
    \mathcal{D}[\hat{\theta}(E), \theta] = \left\vert \alpha \log\left[ \hat{\theta}(E)/ \theta \right] \right\vert^k, 
\end{equation}
where $\alpha$ and $k$ are free-parameters. In the usual setting of least-squares estimation, where $\alpha =1, k=2$, their optimization condition thus means minimization of \emph{mean logarithmic error}, defined below as 

\begin{equation}
    \bar{\epsilon}_{\text{mle}} = \int dE d\theta p(E, \theta) \log^2\left[ \hat{\theta}(E)/ \theta \right]
    \label{eq:mean_log-error}
\end{equation}
\\

Since this minimization is to be done over family of estimators $\hat{\theta}(E)$, or equivalently, over dimensionless estimators $\hat{x}(E) = \alpha \log(k_B \hat{\theta}(E)/\epsilon_0) $, this turns into the familiar  variational calculus problem of optimizing action $\int \mathcal{L}[\hat{x}(E)] dE$, where the effective Lagrangian $\mathcal{L}$ is given by 

\begin{equation}
    \mathcal{L} {=} {\int}d\theta p(E,\theta) \log^2\left[ \frac{\hat{\theta}(E)}{ \theta} \right] {=} {\int} d\theta p(E,\theta) \log^2\left[ \frac{\epsilon_0}{ k_B\theta} e^{\frac{\hat{x}(E)}{\alpha}} \right].
\end{equation}

\noindent The Euler-Lagrange equation now yields
\begin{equation}
    \frac{\partial \mathcal{L}}{\partial\hat{x}} {=}\frac{\partial}{\partial\hat{x}} \left[ {\int} d\theta p(E,\theta) \log^2\left[ \frac{\epsilon_0}{ k_B\theta} e^{\frac{\hat{x}(E)}{\alpha}} \right] \right] {=} \frac{d}{dE}\left[\frac{\partial L}{\partial_E \hat{x}}\right] {=} 0.
\end{equation}
Simplifying after some algebra, the optimal estimator $\hat{\theta}^{*}$ is given by 
\begin{equation}
\frac{k_B\hat{\theta}^{*}}{\epsilon_0} = \exp \left[ \int d\theta p(\theta|E) \log \left(\frac{k_B\theta }{\epsilon_0}\right) \right].
\end{equation}
\noindent The corresponding value of the optimal mean-logarithmic error is given by 

\begin{equation}
    \bar{\epsilon}_{\text{opt}} = \bar{\epsilon}_p - \mathcal{K}, 
\end{equation}
\noindent where the first part is the prior uncertainty $\bar{\epsilon}_p = \int d\theta p(\theta) \log^2 \left(\hat{\theta}_0^{*}/\theta\right)$ with $\hat{\theta}_0^{*}$ being the optimal prior estimate $ \hat{\theta}_0^{*} {=} \epsilon_0 \exp\left[\int d\theta p(\theta) \log(k_B \theta/\epsilon_0)\right]/k_B$ before any measurement readout is obtained. The second part is the reduction of uncertainty from measurement readout $E$, i.e.,
$\mathcal{K} = \int dE p(E) \log^2 \left[ \hat{\theta}^{*}/\hat{\theta}_0^{*} \right]$. 
\\
\\
At this point, it is instructive to consider what would have happened if scale-invariance was not imposed. The figure of merit to optimize over, corresponding to Eq.~\eqref{eq:mean_log-error}, would have been the variance averaged with prior weights over the entire range. We can easily see that this leads back to the familiar Bayesian Van Trees type bound with both prior information and measurement results individually contributing to precision. However, thermometry is peculiarly fortunate in the sense that we know the energy measurement is the only optimal one over any range. What if this is not the case for a generic sensing problem?


\subsubsection{Scale Estimation of General Parameters}


As we saw for the case of thermometry in the preceding subsection, experimental results of energy measurements are the most relevant ones by design. However, for other parameters, the situation can be far more complicated with the best measurement strategy not immediately obvious. Therefore, the question now becomes: \textit{given a generic parameter to be estimated,  measurement data of parameters $\textbf{y}{=}\lbrace y_1,y_2,..,\rbrace$ already known, and another parameter $x$ being measured at each round, how does one construct the estimator for this parameter from experimental data in a scale-independent way?} To express the notion of scale in a concrete way, Rubio in Ref.~\cite{rubio2022quantum} assumes a new vector $\vec{\xi} = \lbrace x,\textbf{y}\rbrace$, and for each component $\xi_i$ of this new vector, imposes the condition that for two values of the parameter $\theta$, say $\theta', \theta''$, all the corresponding indices $\xi_i^{'}, \xi_i^{''}$ satisfy

\begin{equation}
    \frac{\theta'}{\xi_{i}^{'}} = \frac{\theta''}{\xi_{i}^{''}} = \gamma, 
\end{equation}
where $\gamma$ defines the scale of the problem. However the measured quantity $x$ is dimensionless, analogous to the quantity $k_B\theta/\epsilon_0$ in the previous section. Now, since the parameters $\textbf{y}$ are already known beforehand, they can not furnish any information about $\theta$, since otherwise the scale would not be completely unknown before the experiment starts, i.e., the prior $z(\hat{\theta}|\textbf{y})$ corresponding to hypothesis $\hat{\theta}$ for the unknown parameter estimate $\theta$ can be simply written down as $z(\theta)$. To put this into context, for previously considered situations, maximal ignorance imposes the invariance of prior along linear translation of the parameter within the given interval. In contrast, here the invariance of prior is along scaling $\gamma$ of the parameter within $[0,\infty)$, or more restrictedly, any given interval. In statistics literature, for maximum scale-ignorance, the corresponding prior $z(\theta)$ of the hypothesis $\hat{\theta}$ for the unknown parameter is well-known as Jeffrey's prior $z(\theta) \propto 1/\theta$, as we already showed in the previous section. Our first goal is to find the likelihood model $p(x|\theta, \textbf{y})$. From the scaling property, we get $p(x|\theta, \textbf{y}) = p(x|\gamma\theta,\gamma\textbf{y})$. Solving this functional equation restricts the likelihood model to be a function $h(x,\textbf{y}/\theta)$. On the other hand, when the probe state $\rho_{\textbf{y}}(\theta)$ encoding the unknown parameter information is interrogated via a POVM $\Pi_{y}(x)$ to generate data $x$, then this likelihood model is given by 

\begin{equation}
    p(x|\theta, \textbf{y}) = \text{Tr}\left[\Pi_{\textbf{y}}(x) \rho_{\textbf{y}}(\theta)\right] = h(x,\textbf{y}/\theta)
\end{equation}

\noindent Thus, putting everything together, the posterior probability $p(\theta|x,\textbf{y})$ is given via Bayes' theorem as 

\begin{equation}
    p(\theta|x,\textbf{y}) \propto  \frac{h(x,\frac{\textbf{y}}{\theta})}{\theta}
    \end{equation}

\noindent Now, proceeding exactly as before in the thermometry case, we may again choose a deviation function $D(\hat{\theta}/\theta) = \log^2 (\hat{\theta}/\theta)$, we can again define the mean-logarithmic error as the figure of merit to be minimized for a sensing protocol. 

\begin{equation}
    \bar{\epsilon}_{\text{mle}} = \int dx d\theta p(x| \theta, \textbf{y}) \log^2\left( \hat{\theta}/ \theta \right)
\end{equation}

\noindent Using the definition of conditional probability, this can again be written down concretely as a variational calculus problem

\begin{equation}
     \min \bar{\epsilon}_{\text{mle}} = \min_{\Pi_{\textbf{y}} (x), \hat{\theta} (x)} \int dx~d\theta ~ \text{Tr}\left[ \Pi_{\textbf{y}} (x) \rho_{\textbf{y}}(\theta)\right] \log^2\left( \hat{\theta}/ \theta \right)
\end{equation}

The crucial technical difference with thermometry example previously is that we also have to minimize over the measurement strategies $\Pi_{y} (x)$. Applying the Euler-Lagrange equation for matrices, we arrive at the minimum value of mean logarithmic error in the same form as before,

\begin{equation}
    \bar{\epsilon}_{\text{opt}} = \bar{\epsilon}_{p} - \mathcal{K},
\end{equation}
\\
where $\bar{\epsilon}_{p} = \int d\theta p(\theta) \log^{2}\frac{\theta}{\theta_u}$, and $\mathcal{K} = \text{Tr}[\rho_\textbf{y}(\theta)\mathcal{S}_{\textbf{y}}^2]$, with the operators $\mathcal{S}_{\textbf{y}}$ satisfying the Lyapunov equation 
\begin{equation}
    \mathcal{S}_{\textbf{y}} \rho_\textbf{y} + \rho_\textbf{y}  \mathcal{S}_{\textbf{y}}  = 2 \int d\theta~p(\theta)~\log(\theta/\theta_u),
\end{equation}
and the optimal measurement strategy is along eigenvectors $\lbrace|s\rangle\rbrace$ of $\mathcal{S}$ with eigenvalues 
\begin{equation}
    s = \frac{\text{Tr}\left[|s\rangle\langle s|\int d\theta~p(\theta)~\log(\theta/\theta_u) \right]}{\text{Tr}\left[ |s\rangle\langle s| \rho_{\textbf{y}}(\theta) \right]},
\end{equation}
and the optimal estimator is given by $\theta_u \exp\left[\int d\theta p(\theta|s,\textbf{y}) \log(\theta/\theta_u) \right]$. Clearly, this has the same generic form as the thermometry example considered before.  

\section{Frequentist Strategies for Global Quantum Sensing}
\label{sec:Frequentist Strategies for Global Quantum Sensing}
In the previous section, we reviewed the Bayesian approach to global sensing. In this section, we aim to cover the complementary frequentist approach. More specifically, this section reviews a global sensing strategy based on the average uncertainty and its relation to the QFI~\cite{montenegro2021global}. Let $\theta$ represent a single unknown parameter that varies over a sensing interval $[\theta_{\mathrm{min}}, \theta_{\mathrm{max}}]$. Local sensing becomes optimal only when the range of variation $\Delta_{\theta} = \theta_{\mathrm{max}} - \theta_{\mathrm{min}}$ is very \textit{small}, that is the parameter varies over a very narrow interval $\Delta_{\theta}  \rightarrow 0$. In such cases, there is considerable prior knowledge about where the unknown parameter is expected to fluctuate. The ideal measurement in this case can be taken at the central value $\theta_{0} = (\theta_{\mathrm{max}} + \theta_{\mathrm{min}})/2$. Conversely, if the parameter varies over a wider interval $\Delta_{\theta}$, a global sensing strategy is more suitable, as it provides the best sensing approach over such a \textit{larger} sensing interval. Here we note that in specific settings like intermetrometric single-phase estimation based on GHZ states, sometimes one can bypass this optimization of measurement settings even in the absence of a prior \cite{belliardo2020achieving}, as has been experimentally verified with orbital angular-momentum states of light \cite{cimini2023experimental}. However, for general estimation problems, the measurement setting does indeed change. \\

The first general approach that addresses this problem~\cite{montenegro2021global} has been proposed to optimize the probe within a global sensing strategy, aimed at significantly reducing the average uncertainty across the entire sensing interval  $\Delta_{\theta}$. The formulation focuses on minimizing the average uncertainty of the estimation, namely $\int_{\Delta_{\theta}} \mathrm{Var}[\theta] z(\theta) d\theta$, where $z(\theta)$ is the weight representing our information of the unknown parameter $\theta$ across the sensing interval $\Delta_{\theta}$. Note that this merely reflects our initial knowledge about the interval and strictly speaking, is not a prior in the Bayesian sense in that it is not updated to obtain a posterior at each round, reflecting the fact that we are more concerned with the average uncertainty as a figure of merit. Also, this formulation implicitly assumes we know the scale of the parameter and we are interested in the location of the parameter within the given scale. For a more detailed discussion on the latter, we refer the reader to Rubio's work in Ref.~\cite{rubio2022quantum} and Table 1 therein.
By applying the Cram\'{e}r-Rao theorem, it can be shown that this average uncertainty is bounded by
\begin{equation}
g(\boldsymbol{A}) := \int_{\Delta_{\theta}} \frac{z(\theta)}{F_Q(\theta|\boldsymbol{A})}d\theta,\label{eq_global_sensing_gB}
\end{equation}
where $\boldsymbol{A} = (A_1, A_2, \ldots)$ are controlled, known parameters that can be adjusted freely. The optimal probe is determined by minimizing the function $g(\boldsymbol{A})$ with respect to the control parameters $\boldsymbol{A}$:
\begin{equation}
g(\boldsymbol{A}^*) := \min_{\boldsymbol{A}} \left[ g(\boldsymbol{A}) \right].
\end{equation}
Without loss of generality, a uniform weight for $\theta$ within the allowed interval can be considered, i.e. $z(\theta) = (\Delta_{\theta})^{-1}$. In the context of local sensing, minimizing the function $g(\boldsymbol{A})$ is equivalent to maximizing the QFI $F_Q(\theta_{0}|\boldsymbol{A})$ over the uniform distribution $(\Delta_{\theta})^{-1}$.

\subsection{Frequentist Global Quantum Critical Metrology}
To illustrate the relevance of the general formulation for global sensing in Eq.~\eqref{eq_global_sensing_gB}, let us consider its application to strongly correlated quantum many-body probes. Such probes are essential for high-precision measurements, as they can achieve Heisenberg quantum-enhanced precision when are placed at the critical point of a quantum phase transition~\cite{montenegro2024review}. This makes them highly effective for local sensing, where significant prior knowledge about the unknown parameter allows one to place the parameter at their critical point. However, away from criticality, quantum-enhanced precision rapidly reverts to the standard shot-noise limit~\cite{rams2018limits}, with the optimal measurement also depending on the unknown parameter~\cite{paris2009quantum}. These limitations favor a global sensing approach.

\begin{figure}
    \centering
    \includegraphics[width=\linewidth]{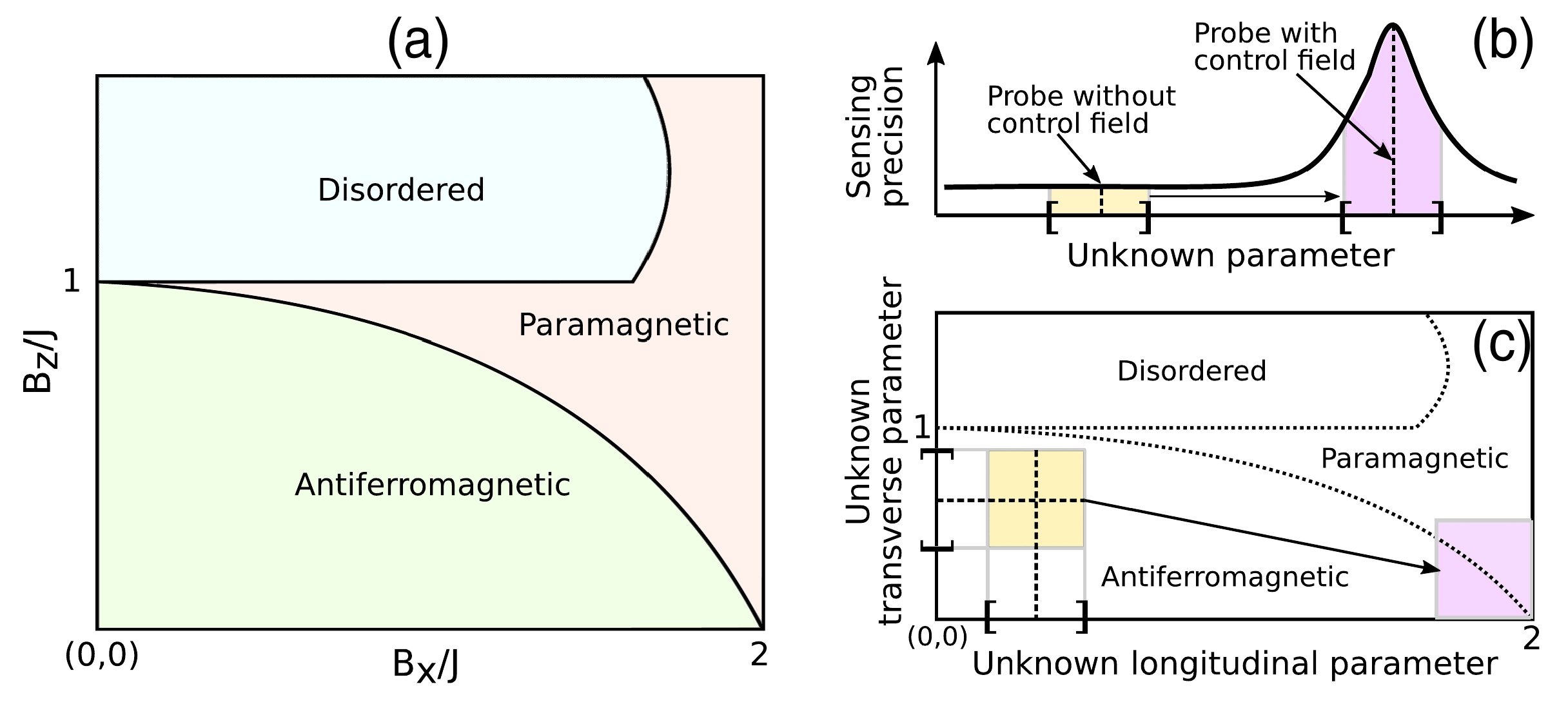}
    \caption{\textbf{Frequentist global spin-chain magnetometry scheme.}(a) Phase diagram for a quantum many-body probe in the presence of longitudinal and transversal magnetic fields. (b) Single-parameter global sensing, where a non-optimal probe transitions to an optimal one by minimizing Eq.~\eqref{eq_global_sensing_gB} over control fields. (c) Multi-parameter global sensing, where the complexity of the phase diagram makes simple offsets insufficient to capture the optimality across the entire phase. Figure taken from Ref.~\cite{montenegro2021global}.}
    \label{fig_sketch_global}
\end{figure}

Fig.~\ref{fig_sketch_global} illustrates the global sensing strategy based on Eq.~\eqref{eq_global_sensing_gB} for both single- and multi-parameter global sensing. In Fig.~\ref{fig_sketch_global}(a), a phase diagram for a quantum many-body probe in the presence of longitudinal and transversal magnetic fields is shown. Since the phase diagram contains critical lines, one might ask where to position the quantum probe for the best estimation of a specific parameter when the unknown parameter varies over a wide sensing range. Fig.~\ref{fig_sketch_global}(b) presents single-parameter global sensing, where a non-optimal probe transitions to an optimal one by minimizing Eq.~\eqref{eq_global_sensing_gB} over control fields. Intuitively, single-parameter global sensing can be seen as an offset from the non-optimal probe to the optimal probe located around the critical point (or, in general, where the QFI is large). However, for multi-parameter global sensing, the situation is more complex, and an offset is insufficient to explain the optimality of the probe across the entire phase diagram, as shown in Fig.~\ref{fig_sketch_global}(c).

For concreteness, consider a quantum many-body probe composed of $L$ interacting spin-$1/2$ particles, governed by an Ising Hamiltonian, which is used to sense a random static magnetic field. The Hamiltonian, assuming periodic boundary conditions, is given by:
\begin{equation}
H = J\sum_{i=1}^{L} \sigma_x^i \sigma_x^{i+1} - \sum_{i=1}^{L}[(B_x + h_x)\sigma_x^i + (B_z + h_z)\sigma_z^i],\label{eq:hamiltonian}
\end{equation}
where $\sigma_\alpha^i$ ($\alpha = x, z$) represents the Pauli operator at site $i$, $J > 0$ is the exchange interaction, $\boldsymbol{B} = (B_x, B_z)$ denotes the tunable control magnetic field, and $\boldsymbol{h} = (h_x, h_z)$ represents the unknown magnetic field to be estimated. While any quantum state can be used to evaluate the QFI, it is known that the ground state $|\Phi\rangle$, for which the phase diagram is shown in Fig.~\ref{fig_sketch_global}(a), enjoys a second-order phase transition~\cite{Bonfim-2019, sachdev1999quantum}, and thus it can readily be used for detecting the unknown field $\boldsymbol{h}$. Indeed, in the specific scenario of single-parameter magnetometry when only transverse field exists $B_x{=}h_x{=}0$, the Hamiltonian simplifies to:
\begin{equation}
H = J\sum_{i=1}^{L} \sigma_x^i \sigma_x^{i+1} - \sum_{i=1}^{L} h_z\sigma_z^i.
\end{equation}
It is possible to diagonalize the above Hamiltonian as follows~\cite{montenegro2021global}:
\begin{equation}
H = \sum_{k>0} \begin{pmatrix}
  \gamma^{\dagger}_{k} &\gamma_{-k}
\end{pmatrix} \begin{pmatrix}
  \epsilon^{+}_{k} & 0\\ 
  0 & \epsilon^{-}_{k}
\end{pmatrix}
\begin{pmatrix}
  \gamma_{k} \\
  \gamma{\dagger}_{-k}.
\end{pmatrix}, \label{eq:sm-FS-hamiltonian}
\end{equation}
where the operators $\{\gamma^{\dagger}_{k},\gamma_{k}\}$ are Bogoliubov operators. The ground state energy of the system is given by 
\begin{equation}
E=-\sum_{k>0}\epsilon_{k}=-\sum_{k>0}\sqrt{(h_z+J\cos(k))^2+J^2\sin^{2}(k)}
\end{equation}
with ground state
\begin{equation}
|\Phi(h_z)\rangle = \Pi_{k}\Big(\cos(\theta_k/2)+\sin(\theta_k/2)c^{\dagger}_{k}c^\dagger_{-k}\Big)|vac\rangle,\label{eq_gs_gs}
\end{equation}
where $\left(\sin(\theta_k/2),\cos(\theta_k/2)\right)=\left( \frac{J\sin k}{\epsilon_k}, \frac{h_z + J\cos k}{\epsilon_k}\right)$. For the above ground state of Eq.~\eqref{eq_gs_gs} which is a function of the parameter $h_z$, it is now straightforward to evaluate the QFI~\cite{paris2009quantum, sidhu2020geometric}.

\begin{figure}
    \centering   \includegraphics[width=\linewidth]{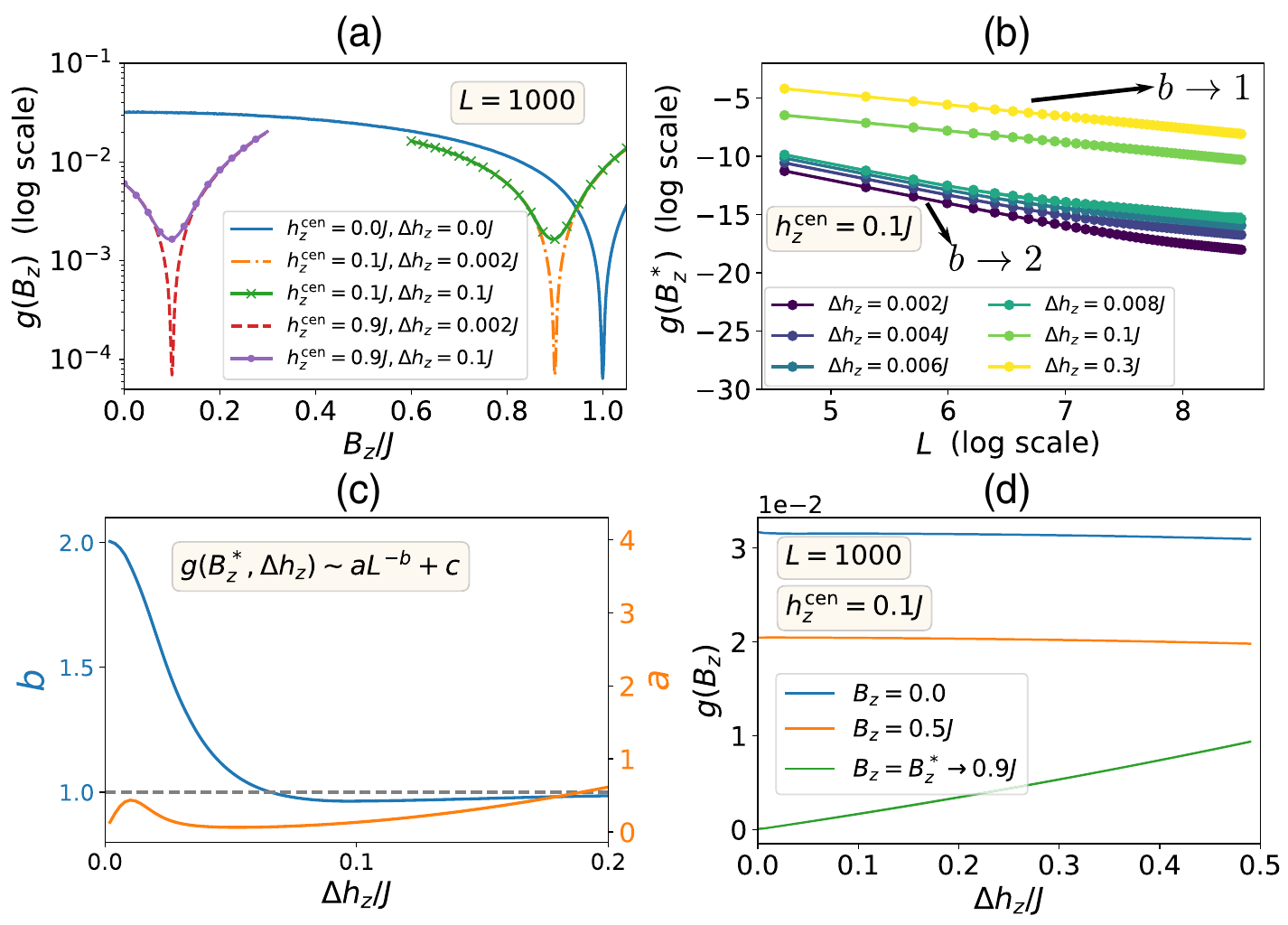}
    \caption{\textbf{ Frequentist global spin-chain magnetometry results.} (a) Average uncertainty $g(B_z)$ as function of $B_z/J$ for different values of interval widths $\Delta_{h_{z}}$ and centers $h^z_0$. The optimal probe can always be found by tuning the control field such that $g(B_z^*) = \mathrm{min}_{B_z}\left[ g(B_z) \right]$. (b) $g(B_z^*)$ (shown by markers) and its corresponding fitting function $aL^{-b}+c$ (solid lines) as a function of $L$ for various choices of $\Delta_{h_{z}}$. (c) Fitting coefficients $a$ and $b$ versus the controlled field $B_z/J$. (d) $g(B_z)$ is plotted as a function of $B_z/J$ for various choices of $B_z$. Figure taken from Ref.~\cite{montenegro2021global}.}
    \label{fig_global_main}
\end{figure}

In Fig.~\ref{fig_global_main}(a), the global sensing performance $g(B_z)$ is plotted as a function of $B_z/J$ for different values of interval widths $\Delta_{h_{z}}$ and centers $h^z_0$. For every $h^z_0$ and $\Delta_{h_{z}}$, the average uncertainty $g(B_z)$ always exhibits a minimum at a specific $B_z{=}B_z^*$, indicating that the probe can be optimized by this choice of control field. Notably, the minimum value of $g(B_z)$ is independent of $h^z_0$ and is solely determined by $\Delta_{h_{z}}$. Conversely, the optimal control field $B_z^*$ is nearly independent of $\Delta_{h_{z}}$ and depends only on $h^z_0$, such that $h^z_0 + B_z^* \approx h^{\mathrm{crit}}$. This implies that the control field tends to shift the probe in its phase diagram, positioning the sensing interval almost symmetrically around the critical point, as illustrated schematically in Fig.~\ref{fig_sketch_global}(b).

Scaling of the sensing precision with respect to a sensing resource is of key importance in sensing stretagies. Within the global sensing strategy outlined above, $g(B_z^*)$ is plotted as a function of the sensing resource $L$ (the system size) for various choices of $\Delta_{h_{z}}$ in Fig.~\ref{fig_global_main}(b). The figure illustrates that for small $\Delta_{h_{z}}$, the average uncertainty scales as $g(B_z^*) \sim 1/L^2$, which is expected for local sensing. Remarkably, as $\Delta_{h_{z}}$ increases, the scaling transitions toward the standard limit, specifically $g(B_z^*) \sim 1/L$. To quantify this transition from quantum-enhanced sensing to the standard limit, we fit $g(B_z^*)$ with a function of the form $aL^{-b}+c$, where $\{a, b, c\}$ are real coefficients. In Fig.~\ref{fig_global_main}(c), the fitting coefficients $a$ and $b$ are plotted as functions of the controlled field $B_z/J$. The figure shows that as the width approaches zero, the expected Heisenberg scaling, $g(B_z^*) \sim {F}_Q^{-1} \sim 1/L^2$, is recovered from the local sensing strategy. Quantum-enhanced sensing is indicated by $b>1$, where the precision exceeds the standard limit of $b=1$. Notably, the region of quantum-enhanced sensing is limited to $\Delta_{h_{z}} \leq 0.07J$, beyond which the standard limit is reinstated. However, it is essential to emphasize that optimizing the probe remains advantageous for sensing, even in the absence of a quantum-enhanced advantage in the scaling. Indeed, it is important not to misunderstand: \textit{losing the Heisenberg limit of precision does not mean losing all quantum-enhanced sensing capabilities}. To check this, in Fig.~\ref{fig_global_main}(d), $g(B_z)$ is plotted as a function of $B_z/J$ for various choices of $B_z$. The figure demonstrates that with the optimal choice of $B_z{=}B_z^*$, the average uncertainty is significantly lower than for non-optimal values of the control field across all values of $\Delta_{h_{z}}$. Specifically, for large $\Delta_{h_{z}}$, while the exponent $b{=}1$ remains fixed, the coefficient $a$ is notably reduced by optimizing $B_z$.

The reviewed global sensing strategy, which minimizes the average uncertainty of Eq.~\eqref{eq_global_sensing_gB}, provides a systematic formalism for addressing global sensing problems through the lens of QFI without prior information about the unknown parameter to be inferred. Moreover, with replacement of the single-number QFI with the full QFI matrix, this technique can be easily extended to multi-parameter scenarios, thus making it a versatile tool for global quantum sensing~\cite{montenegro2021global}. 

\subsubsection{Modular Sensors for Frequentist Global Critical Metrology}

\begin{figure}
    \centering   \includegraphics[width=\linewidth]{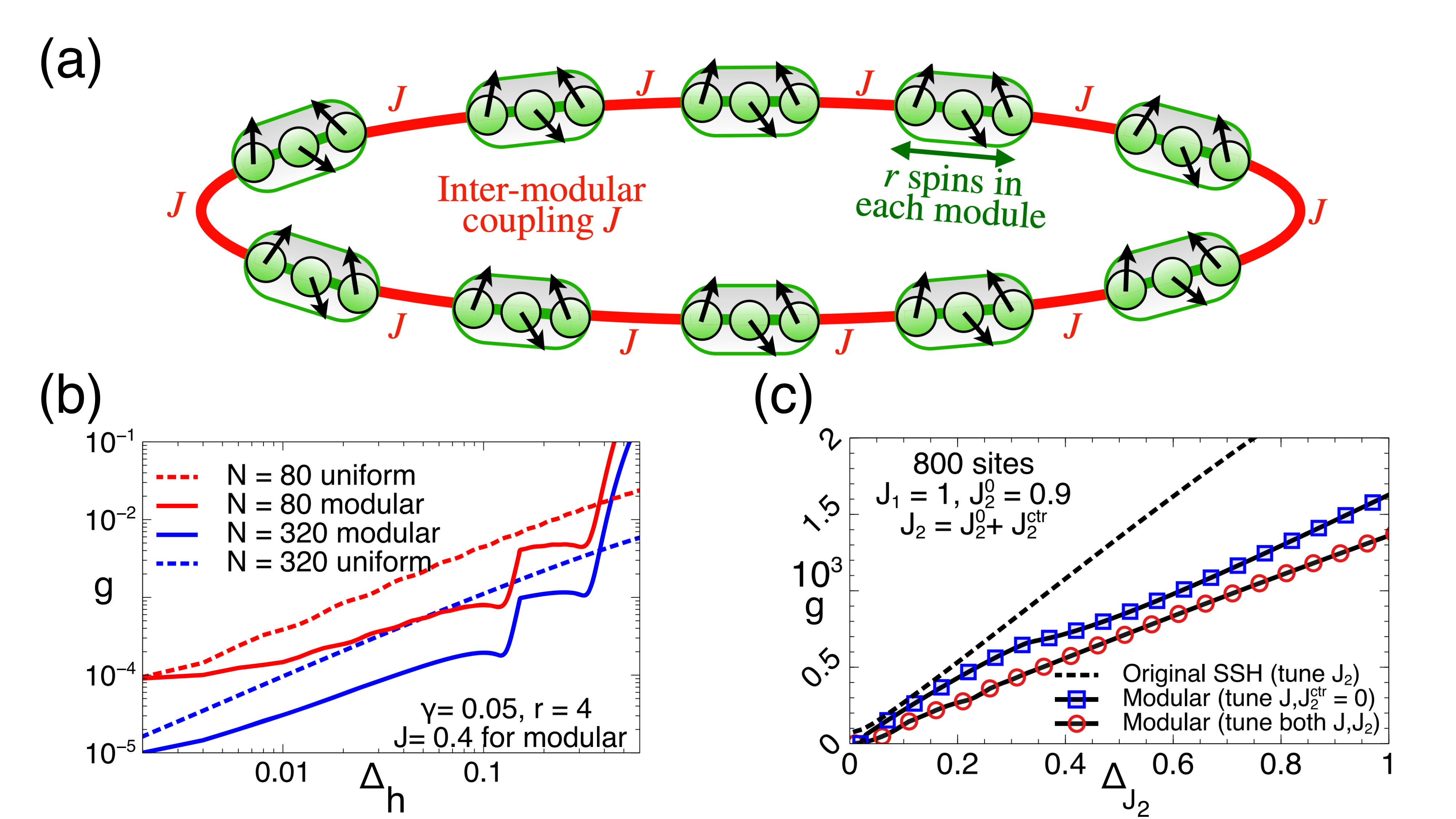}
    \caption{\textbf{ Frequentist global sensing with modular probes.} (a) Schematic of the modular probe. (b) Comparison of global uncertainty optimized with a known control magnetic field in Eq.~\eqref{eq_global_sensing_gB} with width $\Delta_h$ for sensing unknown magnetic field $h$ of anisotropic transverse XY probes (solid lines are modular probes with 4 sites per module, dashed lines are same-sized uniform probes). (c) Comparison of global sensing uncertainty in Eq.~\eqref{eq_global_sensing_gB} with width $\Delta_{J_2}$ for sensing unknown coupling strength $J_2$ with modular SSH chains (solid lines are modular SSH probes with 2 SSH supercells per module, each containing two sites, dashed line is the  conventional SSH probe with identical number of total sites).  Figures taken from Ref.~\cite{mukhopadhyay2024modular}.}
    \label{fig_global_modular}
\end{figure}

This approach above is general enough to be applied to any quantum system, and in fact motivates us towards new methods of sensor engineering. From the example above, it is clear that while the critical region contributes towards enhanced scaling of QFI in the local sensing paradigm, the global sensing scenario incorporating regions deep within either phase destroys this scaling advantage. Thus, a natural idea is to create new phase boundaries associated with enhanced sensitivities, while retaining the nature (i.e., critical exponents) of the criticalities of the original sensor model as much as possible. This was considered in Ref~\cite{mukhopadhyay2024modular} via a modular construction. As depicted in Fig.~\ref{fig_global_modular}(a), the modular construction consists of joining a large number of smaller many-body chains (modules) with adjustable couplings to build a bigger sensor probe. The advantage is that each critical region of the original uniform chain, which is also obtained in the limit of intra-modular couplings equalling the inter-modular couplings, is effectively multiplexed in the composite modular probe. Thus, since these critical regions are known to correspond to quantum-enhanced sensitivity, the effect of modularity is to incorporate these newly created criticalities and effectively widen the performance range of the sensor. From the expression of Eq.~\eqref{eq_global_sensing_gB}, it is easy to see how encompassing all these critical regions enhances the global sensing capability compared to non-modular probes. Ref.~\cite{mukhopadhyay2024modular} demonstrated this through two types of many-body probes undergoing qualitatively different kinds of phase transitions. 

\hfill

\emph{Many body probes undergoing second order phase transitions.---} Let us consider an anisotropic transverse XY probe undergoing second order $U(1)$ continuous symmetry-breaking transition with the Hamiltonian 
    \begin{equation}
        H = \frac{1}{2} \sum_{\langle ij\rangle} J_{ij} \left[ \frac{1+\gamma}{2} \sigma_x^i \sigma_x^j + \frac{1-\gamma}{2} \sigma_y^i \sigma_y^j \right] + h \sum_i \sigma_z^i
    \end{equation}

Like before, the task is to estimate the unknown magnetic field $h$. However, instead of homogeneity assumption for the uniform spin chain considered previously, the nearest-neighbour couplings $J_{ij}$ now come in two varieties. Intra-modular couplings are all $J_{ij} {=}J_0$, while the tunable inter-modular couplings are all $J_{ij} {=} J$. Restricting ourselves to the ferromagnetic sector, the uniform spin chain undergoes a phase transition from ordered (ferromagnetic) to disordered (paramagnetic) phase as the transverse magnetic field strength $h$ overtakes the coupling strength. However, in the modular paradigm, especially for low values of anisotropy, small islands of the disordered phase emerge within the original ordered phase. This leads to multiplication of critical regions, but with the same critical exponents. The number of these islands is equal to the size of each module. Moreover, by varying inter-modular coupling strengths and module sizes, the critical regions can be engineered to be as close or far to each other as desired. See Ref.~\cite{mukhopadhyay2024modular} for more details. For the present purpose, Fig.~\ref{fig_global_modular}(b) demonstrates the power of this approach. The modular probe comfortably outperforms the same-sized uniform probe for typical widths, and in fact performance of smaller modular probes is comparable to far larger uniform probes in the global sensing scenario. 

\hfill

\emph{Many body probes undergoing topological phase transitions.---} Now let us consider the modular version of a Su-Schrieffer-Heeger (SSH) probe undergoing  discrete chiral symmetry-breaking topological phase transition with the Hamiltonian 
\begin{equation}
    H = \sum_{i,\mu} \left[ J_{i} c_{i,\mu}^{\dagger}c_{i+1,\mu} +  J_{\mu}c_{r,\mu}^{\dagger}c_{1,\mu+1}\right],
\end{equation}
where the Roman indices denote site indices within each module, and the Greek indices denote module indices. There are $r$-sites per module as before. For the conventional SSH chain with each module consisting of only two sites, $J_i {=} J$ and $J_\mu {=} J'$, the system undergoes a topological phase transition from topologically trivial bulk to topologically active bulk at $J = J'$. This transition is manifested in emerging edge-localized states, whose sensing capability have been studied in Ref.~\cite{free2022sarkar}. In its simplest form, modularization leads one to consider four sites per module, with $J_i = \lbrace J_1, J_2, J_1 \rbrace$ respectively, and tunable intermodular coupling $J_\mu = J$. By setting $J{=} J_2$, one can recover the original SSH chain. One can without loss of generality set $J_1{=}1$, and the sensing task is to estimate the unknown intramodular coupling $J_2$ using the tunable intermodular coupling $J$. In this case, Ref~\cite{mukhopadhyay2024modular} showed that there are now two topological phase transitions at $J{=}J_2$ and $J{=}1/J_2$. For the many-body ground state, Fig.~\ref{fig_global_modular}(c) confirms that the same intuition leads to far better global sensing capability compared to similarly sized conventional SSH chains. This phenomenon of multiplexing of critical regions has also been reported in 2D SSH models with various tilings in Ref~\cite{mukhopadhyay2024modular}.

\subsection{Frequentist Global Quantum Thermometry}

Let us now consider equilibrium thermometry in the frequentist global estimation picture with the bound obtained in Eq.~\eqref{eq_global_sensing_gB}. \textcolor{black}{To set the notation, we assume the probe is fully thermalized at a temperature $T$, i.e., if the probe Hamiltonian is $H$, then the quantum state of the probe is the canonical thermal Gibbs state given by the density matrix $\rho_{\text{thermal}}= e^{-H/k_B T}/ \text{Tr}[e^{-H/k_B T}]$, where $k_B$ is the Boltzmann constant}. Since thermal states are always diagonal in the energy basis, unlike the previous case there is no ambiguity over the optimal measurement basis. Instead, here is the challenge is to engineer the energy levels $\lbrace E_n \rbrace$ of an $N$-level probe Hamiltonian to ensure maximum precision. In the local case, this problem was treated in Ref.~\cite{correa2015individual}, and yielded the following solution for the optimal probe - 

\begin{enumerate}
    \item The optimal probe has a single ground state and an $N-1$ fold degenerate excited state. That is, the optimal qudit probe behaves essentially like a two-level system. 
    
    \item For a temperature $T$, the optimal energy gap $\epsilon$ is given by the solution of the transcendental equation 
    \begin{equation}
        e^x = (N-1)\frac{x+2}{x-2},
    \end{equation}
    where $x = \epsilon/k_B T$ is the dimensionless energy gap. 
\end{enumerate}

\begin{figure}
    \centering   \includegraphics[width=\linewidth]{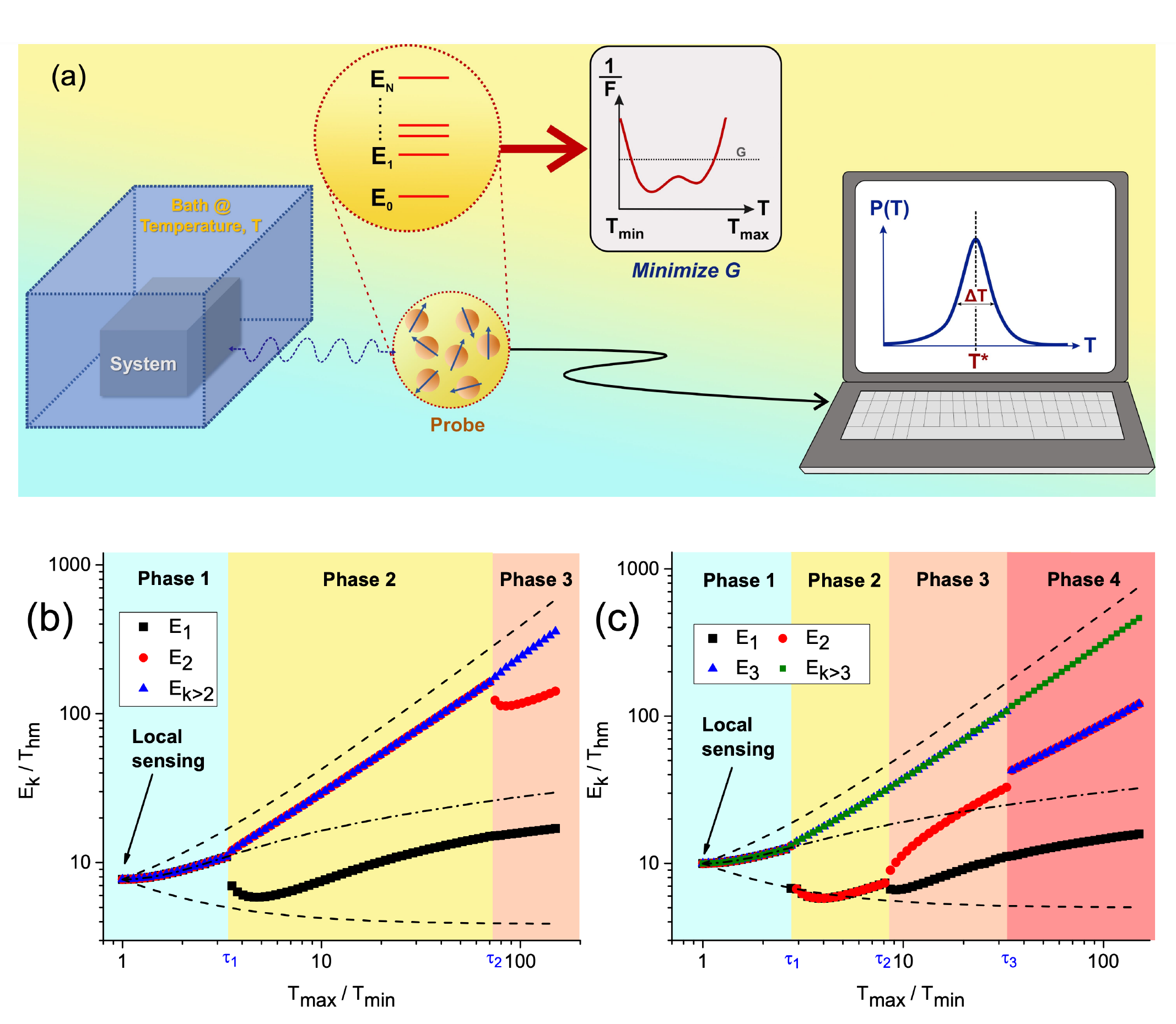}
    \caption{\textbf{Frequentist global quantum thermometry.} (a) Schematic of proposal. A probe weakly interacts with the system and thus reaches an equilibrium at the system temperature $T \in [T_{\text{min}},T_{\text{max}}]$ followed by Energy measurements on the probe. The global thermometry measure $g$ is defined as per Eq. (\ref{eq_global_sensing_gB}). The optimal energies against relative width $T_\text{max}/T_\text{min}$ for a general $N$-level system is obtained for: (b) $N = 16$, (c) $N = 64$. The energies are scaled with the characteristic temperature $T_{hm} = \frac{T_\text{min} T_\text{max}}{T_\text{min} + T_\text{max}}$. The leftmost data point corresponds to the case of the locally optimal thermometer, dash-dotted line shows the optimal energy gap for an effective two-level system. Figure taken from Ref.~\cite{mok2021optimal}.}
	\label{fig:qudit_bifurcation}
\end{figure}

\begin{figure}
    \centering   \includegraphics[width=\linewidth]{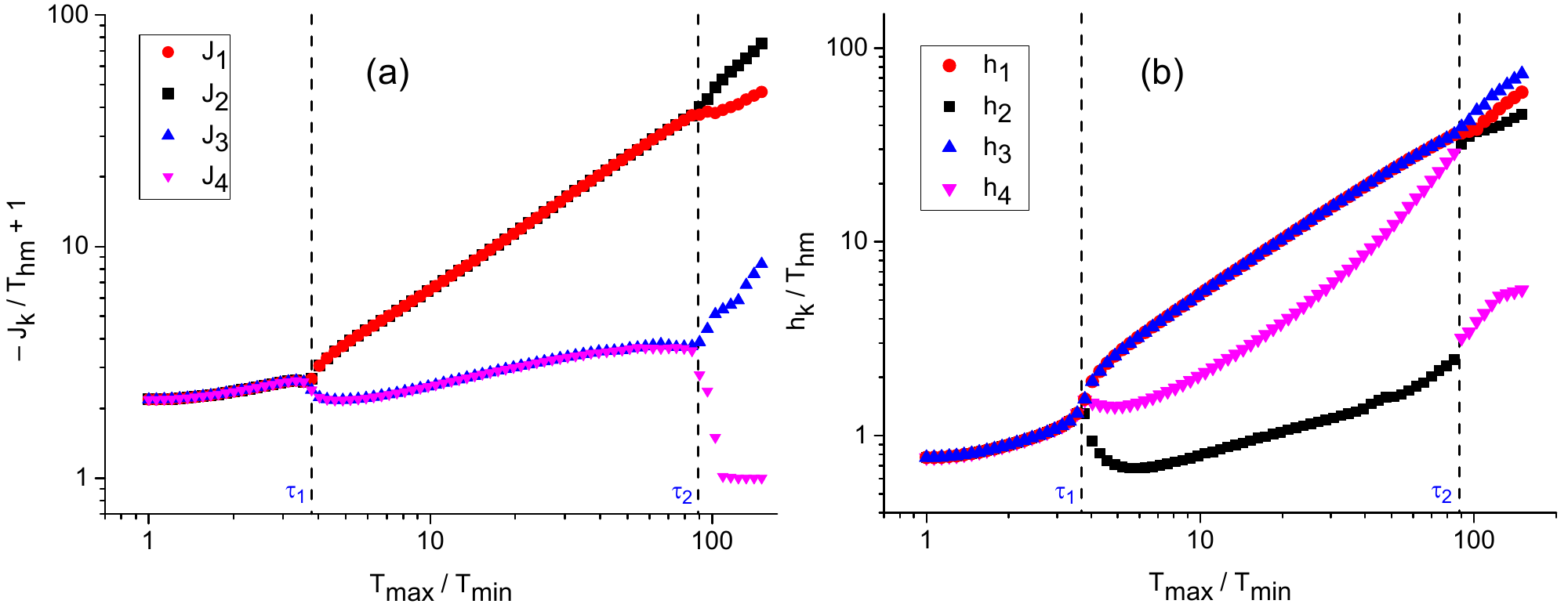}
    \caption{\textbf{Optimal probe for $ \text{XYZ}$-Hamiltonian.} Optimal parameters against temperature ratio $T_\text{max}/T_\text{min}$ for an optimized $H_{\text{XYZ}}$ (reduced to an Ising model) for $n = 4$ spins. (a) ZZ-coupling strengths $J_k$. For illustration purposes, we plot $-J_k$ shifted positively by 1 unit. (b) External magnetic fields. The parameters are scaled with the characteristic temperature $T_{hm} = \frac{T_\text{min} T_\text{max}}{T_\text{min} + T_\text{max}}$. The dashed lines indicate the onset of the splitting of optimal parameters. Figure taken from Ref.~\cite{mok2021optimal}.}
	\label{fig:xyz_bifurcation}
\end{figure}

In the global case, let us assume the temperature $T$ is known to lie within the interval $T \in [T_{\min}, T_{\max}]$ with an uniform weight. By minimizing the figure of merit given in Eq.~(\ref{eq_global_sensing_gB}) with respect to energy levels one can obtain the optimal probe design for global thermometer, see Fig.~\ref{fig:qudit_bifurcation}(a) for the schematic of the procedure. Ref.~\cite{mok2021optimal} demonstrated that  as the relative width of the interval grows, the optimal energy configuration splits from an effective two-level probe into an effective three-level probe. As the interval gets even broader, the optimal probe is an effective four-level probe, and so on. As shown in Fig.~\ref{fig:qudit_bifurcation}(b) and~\ref{fig:qudit_bifurcation}(c), these splittings occur in discrete steps akin to phase transitions (note that the ground energy is assumed to be $E_0{=}0$). Of course, the resulting Hamiltonians are still highly artificial, therefore a practical question emerges: Do we see a similar splitting with respect to coupling parameters of simple spin-chain probes? For local estimation, one can use symmetry to argue that translationally-invariant, i.e., uniform probes are optimal in this scenario. However, for global estimation,  Ref.~\cite{mok2021optimal} indeed answers this question in the affirmative as non-uniform probes become optimal as the width of the interval increases. As a concrete illustration, the 1D (non-uniform) Heisenberg $\text{XYZ}$-model with periodic boundary conditions was considered with the following Hamiltonian.
\begin{equation}
\begin{split}
H_{\text{XYZ}} &= \sum_{i=1}^{n} (J_x^{i} \sigma_x^{i} \sigma_x^{i+1} + J_y^{i} \sigma_y^{i} \sigma_y^{i+1} + J_z^{i} \sigma_z^{i} \sigma_z^{i+1}) \\
&+ \sum_{i=1}^{n} (h_x^{i} \sigma_x^{i} + h_y^{i} \sigma_y^{i} + h_z^{i} \sigma_z^{i})
\end{split}
\end{equation}
with the presence of local external magnetic fields in all three directions. Interestingly, by performing the optimization of Eq.(\ref{eq_global_sensing_gB}) with respect to all the couplings and the magnetic fields, one obtains that $J_x^{i}{=}J_y^{i}{=}0$ and $h_x^{i}{=}h_y^{i}{=}0$. In other words, the optimal probe becomes a classical Ising system with longitudinal fields. In Fig.~\ref{fig:xyz_bifurcation}(a), the optimal values of the couplings $J_z^{i}$ (for convenience normalized $J_z^{i}$ are plotted) is depicted as a function of $T_{\max}/T_{\min}$. In Fig.~\ref{fig:xyz_bifurcation}(b), the corresponding optimal values of $h_z^{i}$ is plotted as a function of $T_{\max}/T_{\min}$. As both figures show by enlarging the temperature interval, more distinct couplings and magnetic fields are required for optimal global sensing.   

In general, having nonuniform coupling along $x$, $y$ and $z$ directions might be challenging in some physical setups, e.g. quantum dot arrays. In such systems, isotropic couplings naturally arise as $J_x^{i} = J_y^{i} = J_z^{i} = J^{i}$. The only degrees of freedom in such systems is to tune the coupling between neighboring particles. Let us consider isotropic non-uniform Heisenberg XXX Hamiltonian:
	\begin{equation}
	\begin{split}
H_{\text{XXX}} &= \sum_{i=1}^{n} J^{i} (\sigma_x^{i} \sigma_x^{i+1} + \sigma_y^{i} \sigma_y^{i+1} + \sigma_z^{i} \sigma_z^{i+1}) \\
&+ \sum_{i=1}^{n} (h_x^{i} \sigma_x^{i} + h_y^{i} \sigma_y^{i} + h_z^{i} \sigma_z^{i})
	\end{split}
	\end{equation}
In practice, it is difficult to engineer exchange couplings with varying signs. Therefore, we impose the constraint that the spin chain probe must either be antiferromagnetic (all $J > 0$) or ferromagnetic (all $J < 0$), while the signs of the local magnetic fields can be arbitrarily varied. By numerically optimizing the Eq.~(\ref{eq_global_sensing_gB}) with respect to the couplings $J^i$ and magnetic fields $\{h_x^{i},h_y^{i},h_z^{i}\}$, one gets $h_x^{i}{=}h_y^{i}{=}h_z^{i}{=}0$. The optimal couplings $J_i$, however, show interesting behavior and only alternating couplings becomes nonzero, i.e. $J_i{=}0$ for  $i$ being even. In other words, the optimal XXX probe is combination of dimers. Therefore, for a chain of $n=6$, there are only three nonzero optimal couplings. The value of these couplings are initially identical and as $T_{\max}/T_{\min}$ enlarges one by one take seperate values, see Ref.~\cite{mok2021optimal} for details.

\subsection{Saturable Global Sensing in Frequentist Picture}
We saw an example of a frequentist global sensing bound in Eq.~\eqref{eq_global_sensing_gB}, where the right-hand side of Eq.~\eqref{eq_global_sensing_gB} was expressed in terms of QFI $F_Q$. For tasks like equilibrium thermometry~\cite{mok2021optimal}, where the optimal measurement basis remains unchanged throughout any interval, this is perfectly saturable. However, generally the optimal measurement basis can rotate significantly through the finite interval typical of global sensing. Bayesian and other adaptive strategies play around this problem by essentially mapping the global sensing problem to a local one through repeated updating, as we saw above.  Moreover, especially as Hayashi's works~\cite{hayashi2006parallel,hayashi2011comparison,hayashi2018resolving} show, Heisenberg scaling in this sense may not be generally attainable even with adaptive strategies and the best global estimator may in fact differ significantly from the optimal local estimator even when the width is very small. This naturally leads to the so-called \emph{large-deviation analysis}~\cite{hayashi2002two}, where we ask the following question: \emph{What is the worst case probability, i.e., the probability $p(|\hat{\theta} - \theta| > c)$ with which the error in estimation exceeds a certain threshold $c$?} In the context of optical estimation of an unknown phase - Hayashi, Liu, and Yuan \cite{hayashi2022global} pointed out that to preserve Heisenberg scaling globally, the threshold $c$ must scale as $c\sim O(n^{-1})$ with the number of channels $n$ being used counting as the relevant resource. 
Armed with this insight, they proved that global Heisenberg scaling is attainable even in a noisy setting iff the strength $p$ of the phase-damping noise also scales as $c\sim O(n^{-1})$. However, the measurement strategy to obtain this scaling is highly adaptive in nature. Thus, one naturally asks: Can we define a saturable bound with a single measurement setup for the global sensing problem ? \\

In a frequentist setting, Ref.~\cite{mukhopadhyay2024saturable} indeed takes this approach by considering the CFI with respect to the same measurement strategy throughout the interval, and then performing an optimization of the corresponding figure of merit over the measurement strategy itself. That is, for the same setting as in Ref.~\cite{montenegro2021global}, i.e., of minimizing the average variance of estimator of parameter $\theta$ over the entire range $\theta \in [\theta_{\text{min}},\theta_{\text{max}}]$ known with a weight $z(\theta)$, the lower bound $\mathcal{G}$, termed as \emph{saturable global uncertainty}  in Ref.~\cite{mukhopadhyay2024saturable}, is given as

\begin{equation}
      \int_{\theta_{\text{min}}}^{\theta_{\text{max}}} \!\!\!\!\!\!\!d\theta z(\theta) \text{Var}[\hat{\theta} - \theta ]  {\geq} \min_{\Pi}\int_{\theta_{\text{min}}}^{\theta_{\text{max}}} \!\!\!\!\!\!\!  d\theta \frac{z(\theta)}{F_{c} \left(\theta, \Pi \right)} {=} \mathcal{G}(\theta_{\text{min}}, \theta_{\text{max}}).
    \label{eq:avg_uncertainty}
\end{equation}

The width of the interval is denoted, as before, by $\Delta_\theta {=}\theta_{\text{max}} {-}\theta_{\text{min}}$. Moreover, $\int_{\theta_{\text{min}}}^{\theta_{\text{max}}} ~ d\theta \frac{z(\theta)}{F_{c} \left(\theta, \Pi \right)}$ without optimization for measurement basis is denoted as $\tilde{\mathcal{G}}(\Pi,\theta_{\text{min}}, \theta_{\text{max}})$, such that $\min_{\Pi} \tilde{\mathcal{G}}(\Pi,\theta_{\text{min}}, \theta_{\text{max}}) = \mathcal{G} (\theta_{\text{min}}, \theta_{\text{max}})$. Note that this bound is saturable by construction through a single measurement strategy. The price to pay is that the optimal strategy is hard to derive analytically, since we have to calculate CFI with respect to all the measurements individually, a notoriously difficult problem for large systems, which the calculation of QFI $F_Q$ sidesteps through its fidelity susceptibility interpretation. However, for relatively simpler systems, this is indeed doable in a straightforward manner.

\emph{Gaussian Example.---} A Gaussian state $\rho(\theta)$ encoding information about an unknown parameter $\theta$ can be expressed completely in terms of its first two moments, namely the displacement vector $d(\theta)$ and the covariance matrix $\sigma(\theta)$. Likewise, Gaussian measurements $\Pi$ are also completely represented by their displacement vector $d_m$ and covariance matrix $\sigma_m$. In this case, the expression of CFI $F_c$ with respect to parameter $\theta$ and any Gaussian measurement $\Pi$ is known analytically.

\begin{equation}
    F_c (\theta) = \partial_\theta d^T (\sigma + \sigma_m)^{-1}\partial_\theta d + \frac{1}{2} \textrm{Tr} [(\sigma + \sigma_m)^{-1} \partial_\theta \sigma (\sigma + \sigma_m)^{-1} \partial_\theta \sigma ] 
    \label{eq:fc_gaussian}
\end{equation}

Note that this expression does not involve the displacement $d_m$ of the measurement apparatus, hence only the rotation and squeezing angles of the measurement apparatus have to be considered, confining us to the so-called general-dyne measurement paradigm. This expression of CFI above can be used in theory to optimize over all parameters of $\sigma_m$ to obtain saturable global uncertainty $\mathcal{G}$. For $N$-mode Gaussian probes, there are at most $2N$ parameters ($N$ rotation and $N$ squeezing) to optimize over. For the simplest case of single-mode Gaussian estimation, Ref.~\cite{mukhopadhyay2024saturable} works out two examples - Gaussian thermometry and Gaussian phase estimation. Notice that the Gaussian thermometry problem is different in the sense that the universally optimal measurement basis, the Fock basis, is highly non-Gaussian, and hence not accessible in linear optical, i.e., Gaussian setups. Moreover, one needs to resolve, in theory, each of the countably infinite energy levels perfectly, to perform this optimal measurement. Thus, there indeed is a non-trivial optimization to be performed over Gaussian general-dyne measurements. In both cases, a fascinating general trend emerges. As as one widens the sensing interval, the optimal Gaussian measurement strategy shifts from homodyne to heterodyne. \textcolor{black}{Ref.~\cite{mukhopadhyay2024saturable} shows that this shift is an abrupt flip in the case of Gaussian thermometry.} It is to be noted that this flip occurs for local estimation as well~\cite{cenni2022thermometry} when the temperature $T$ to be estimated crosses a certain threshold $T^{*}$, but the global estimation scheme shifts this threshold temperature $T^{*}$ by an amount depending on the width of the interval. In contrast, the case of Gaussian phase estimation is far more dramatic. For local estimation, homodyne is the optimal strategy for every parameter value~\cite{oh2019optimal}, but with increasing width of the interval, as shown in Ref.~\cite{mukhopadhyay2024saturable}, the optimal measurement logarithmically approaches the  heterodyne in the asymptotic limit. 

\section{Conclusion and Outlook}
\label{sec:Outlook}
Quantum sensing has emerged as a key pillar of upcoming quantum technologies. While most works on quantum sensing assume the local paradigm, the operationally important problem of global quantum sensing has recently begun to receive significant attention. In fact, in recent years, several competing approaches to global quantum sensing have been proposed. In this article, we have reviewed these approaches, dividing them into two main classes: those based on Bayesian updating and those based on a frequentist framework. In the Bayesian approach, the precision bound is achievable; however, this requires adaptive updating of measurement settings. In the frequentist approach, the measurements do not need to be updated from round to round.
In addition, in this latter approach, one can choose between a loose but easy-to-compute, and a saturable but generally harder-to-compute precision bound. The saturable bound has been demonstrated in a simple one-mode Gaussian setting.

Global quantum estimation is a relatively new field, and answers to many critical questions are yet unknown. Below we anticipate only a few of many open directions of further work. Firstly, multiparameter estimation in the global sensing paradigm opens up intriguing possibilities. Mihailescu \emph{et al} showed recently in Ref~\cite{mihailescu2024uncertain} in the context of many-body systems that simultaneous measurement of more than one parameter can lead to the quantum Fisher information matrix becoming singular, and generally a trade-off exists between attainable sensitivity and their robustness to perturbations which can be viewed as the parameter being known only within a finite width. Sequential strategies similar to Ref.~\cite{burgarth2015quantum,montenegro2022sequential,yang2023extractable,yang2024sequential} remain unexplored for global quantum estimation, even though many important results in sequential inference is already known in the statistics community \cite{wald2004sequential}. Another relatively unexplored area is the use of global sensing in distributed scenarios, See Ref.~\cite{zhang2021distributed} and references therein, where a mesh of many sensors serve to build a more complete image of the parameter being sensed. In another direction, motivated by the successful application of optimal control theory to local quantum estimation~\cite{pang2017optimal}, the implementation and calibration of control pulses~\cite{qian2022fast} tailored to the global sensing scenario promise practically useful results.

\bigskip

\emph{Acknowledgments.--}  We are grateful for discussions with George Mihailescu, Jesus Rubio, Matteo Paris, and Daniel Braun. We also acknowledge support from the National Natural Science Foundation of China (Grants No. 12050410253, No. W2432005, No. 92065115, No. 12374482, and No. 12274059), and the Ministry of Science and Technology of China (Grant No. QNJ2021167001L).

\bibliography{global_sensing_review}


\end{document}